\documentclass[floats,floatfix,showpacs,amssymb,prl,superscriptaddress,nofootinbib, aps,twocolumn]{revtex4-1}
\usepackage{graphicx, epsfig, amssymb} 
\usepackage{amsmath, amsfonts}
\usepackage{bm} 

\usepackage[final,breaklinks]{hyperref}
\usepackage[caption=false]{subfig}
\usepackage[usenames]{color}

\def\be{\begin{equation}}
\def\ee{\end{equation}}
\def\beq{\begin{eqnarray}}
\def\eeq{\end{eqnarray}}

\newcommand{\bea}{\begin{eqnarray}}
\newcommand{\eea}{\end{eqnarray}}
\newcommand{\ben}{\begin{enumerate}}
\newcommand{\een}{\end{enumerate}}
\newcommand{\bi}{\begin{itemize}}
\newcommand{\ei}{\end{itemize}}

\newcommand{\nn}{\nonumber}

\begin{document}

\title{Surface singularities in Eddington-inspired Born-Infeld gravity
}

\author{Paolo Pani}
\affiliation{CENTRA, Departamento de F\'{\i}sica, Instituto Superior
T\'ecnico, Universidade T\'ecnica de Lisboa - UTL, Avenida~Rovisco Pais 1, 1049
Lisboa, Portugal}
\author{Thomas P. Sotiriou} 
\affiliation{SISSA, Via Bonomea 265, 34136, Trieste, 
Italy {\rm and} INFN, Sezione di Trieste, Italy.}

\begin{abstract}
Eddington-inspired Born-Infeld gravity was recently proposed as an alternative to general relativity that offers a resolution of spacetime singularities. The theory differs from Einstein's gravity only inside matter due to nondynamical degrees of freedom, and it is compatible with all current observations. We show that the theory is reminiscent of Palatini $f({\cal R})$ gravity and that it shares the same pathologies, such as curvature singularities at the surface of polytropic stars and unacceptable Newtonian limit. This casts serious doubts on its viability. 
\end{abstract}

\maketitle
Recent years have witnessed a proliferation of alternative theories of gravity, motivated by long-standing puzzles in general relativity~\cite{Clifton:2011jh}. 
It is pertinent to develop theoretical benchmarks that will allow us to select the physically relevant candidates, and to subsequently use observational constraints in order to single out the few theories that are actually viable alternatives.

One of the major riddles of general relativity is that it predicts the appearance of spacetime singularities originating from regular initial data, e.g. in the gravitational collapse of massive stars and in the early universe.
In order to resolve these singularities, an appealing proposal for a modified theory of gravity, the so-called Eddington-inspired Born-Infeld (EiBI) theory, was recently put forward in Ref.~\cite{Banados:2010ix} and it has been subject to scrutiny in a number of works~\cite{Pani:2011mg,Casanellas:2011kf,Avelino:2012ge,Pani:2012qb,Delsate:2012ky,Liu:2012rc,EscamillaRivera:2012vz,Avelino:2012ue,Avelino:2012qe,Sham:2012qi,Cho:2012vg}.
EiBI gravity is equivalent to general relativity in vacuum and does not propagate any degree of freedom other than a massless graviton. On the other hand, the theory introduces nonlinear couplings to the matter fields~\cite{Pani:2012qb,Delsate:2012ky}, which resolve at least some of the singularities appearing in Einstein's theory.

The Big-Bang singularity in early cosmology is replaced by a freezing or a bouncing behavior of the cosmological scale factor, depending on the extra EiBI parameter~\cite{Banados:2010ix}. The gravitational collapse of noninteracting particles does not lead to singular states in the nonrelativistic limit~\cite{Pani:2011mg,Pani:2012qb}. 
A tensor instability of the homogeneous and isotropic universe was found in Ref.~\cite{EscamillaRivera:2012vz} and EiBI gravity has been also studied as an alternative to the inflation paradigm~\cite{Avelino:2012ue}. Possible constraints on the theory have been considered using solar models~\cite{Casanellas:2011kf} and cosmological observations~\cite{Avelino:2012ge,Cho:2012vg} (see also Ref.~\cite{DeFelice:2012hq}). However, a degeneracy between EiBI corrections and different matter configurations~\cite{Delsate:2012ky} makes it difficult to put observational constraints without independent knowledge of the matter content of the theory.

Previous literature on EiBI gravity mostly focused on phenomenological aspects of the theory; a more detailed study on its dynamics and on the structure of its field equations has not been performed yet. Here, we argue that the field equations of EiBI gravity have a peculiar differential structure which is similar to that of Palatini $f({\cal R})$ gravity~\cite{Sotiriou:2008rp} and, as a result, they exhibit the same pathologies with the latter (see e.g. Refs.~\cite{Flanagan:2003rb,Barausse:2007pn,Barausse:2007ys,Barausse:2008nm} and the review~\cite{Sotiriou:2008rp}) and with theories where matter is coupled to the Ricci scalar, which also have similar characteristics~\cite{Sotiriou:2008dh}. 
Although these theories provide an appealing (and in fact similar~\cite{Olmo:2009xs}) early-time cosmology, such pathologies cast serious doubts on their viability.

EiBI gravity is described by the following action~\cite{Banados:2010ix}
\begin{eqnarray}
S &=& \frac{1}{4\pi G\kappa}\int d^4x\Big(\sqrt{|\det\left(g_{ab} + \kappa {\cal R}_{(ab)}\right)|}  \nn\\
&&\qquad\qquad- (1+\kappa \Lambda) \sqrt{g}\Big)+S_M\left[g_{ab},\Psi_M\right]\,,\label{action}
\end{eqnarray}
where $S_M\left[g_{ab},\Psi_M\right]$ is the matter action, $\Psi_M$ generically denotes any matter field, ${\cal R}_{ab}$ is the Ricci tensor built from the connection $\Gamma_{ab}^c$, $g=|\det(g_{ab})|$, $\Lambda$ turns out to be the cosmological constant and $\kappa$ is the extra EiBI parameter which has dimensions of length squared. Round (square) brackets denote (anti)-symmetrization.

In the metric approach the field equations contain ghosts, which must be eliminated by adding extra terms to the action~\cite{Deser:1998rj,Vollick:2003qp}.
Thus, EiBI gravity is naturally based on the Palatini formulation, {\em i.e.}~the connection $\Gamma_{ab}^c$ is considered as an independent field. 
The original EiBI proposal is based on some crucial \emph{assumptions}, namely: (i)~the matter action is independent from $\Gamma^c_{ab}$; (ii) the connection is symmetric, $\Gamma_{ab}^c=\Gamma_{ba}^c$; (iii) only ${\cal R}_{(ab)}$ appears in Eq.~\eqref{action} and not ${\cal R}_{ab}$. This last assumption is often implicit. That is, it is common in the literature to have ${\cal R}_{ab}$ appearing in Eq.~\eqref{action}, even though any subsequent calculation is based on the implicit assumption ${\cal R}_{[ab]}=0$ (or on the non-standard definition ${\cal R}_{ab}\equiv {\cal R}_{(ab)}$). This would be an extra constraint as, even for a symmetric connection, ${\cal R}_{[ab]}=-\partial_{[b}\Gamma^l_{a]l}$ and it does not vanish generically. This extra constraint is not required provided that only the symmetric part of the Ricci is used in the action. See Ref.~\cite{Vitagliano:2010pq} for a similar discussion for generalized Palatini gravity.

In principle, assumptions (i), (ii) and (iii) are not required and, together with the action~\eqref{action}, they define a particular version of EiBI gravity. Relaxing (i)--(iii) would lead to a metric-affine version of the theory, similarly in spirit to the case of metric-affine $f({\cal R})$ theories~\cite{Sotiriou:2006qn}, but very different from the original theory. We will explore this possibility in a separate publication. Here we shall rely on the assumptions above,  as in the original proposal~\cite{Banados:2010ix}.

We start by expanding the action (\ref{action})
at second order in $\kappa$. This yields
\begin{eqnarray}
S&=&\frac{1}{8\pi G}\int d^4x\sqrt{g}\left[{\cal R}-2\Lambda+\frac{\kappa}{4}\left({\cal R}^2-2 {\cal R}_{(ab)}{\cal R}^{(ab)}\right)\right]\nn\\
&&+S_M\left[g_{ab},\Psi_M\right]+{\cal O}(\kappa^2) \label{action2}
\end{eqnarray}
where ${\cal R}=g^{ab}{\cal R}_{ab}$.
For simplicity, we shall use units such that $8\pi G=1$. 
When $\kappa=0$, EiBI gravity reduces to the Palatini formulation of general relativity with a cosmological constant. As is well known, in this case the field equations impose that the connection is the Levi-Civita one and the theory reduces to Einstein's gravity. However, at first order in $\kappa$, quadratic corrections in the curvature tensor built from the independent connection appear in the action~\eqref{action2}. The Palatini formulation guarantees that, despite these extra terms, no higher derivatives of the metric field would appear in the field equations. Note also that, when expanded order by order in $\kappa$, the action~\eqref{action} takes the form of a specific Palatini $f({\cal R},{\cal R}_{ab})$ theory~\cite{Vitagliano:2010pq,Olmo:2009xy}.

We now return to action (\ref{action}). Independent variation with respect to the metric and the connection yields
\begin{eqnarray}
&& \sqrt{q}q^{ab}=\sqrt{g}\left((1+\kappa\Lambda) g^{ab}-\kappa T^{ab}\right) \label{eqALG}\,,\\
 && \tilde{\nabla}_c[\sqrt{q} q^{(ab)}]-\tilde{\nabla}_l[\sqrt{q} q^{(al}]\delta^{b)}_c=0\,,\label{varGamma}
\end{eqnarray}
where we have defined $q_{ab}\equiv g_{ab}+\kappa {\cal R}_{(ab)}$
and $\tilde{\nabla}_a$ is the covariant derivative defined with $\Gamma_{ab}^c$, $T^{ab}\equiv (g)^{-1/2}\delta S_M/\delta g_{ab}$ is the standard stress-energy tensor, whose indices are raised and lowered by $g_{ab}$, whereas $q^{ab}$ is the inverse of $q_{ab}$. After some manipulations, Eq.~(\ref{varGamma}) takes the form
\begin{equation}
 \Gamma_{ab}^c=\frac{1}{2}q^{cd}\left(\partial_a q_{bd}+\partial_b q_{ad}-\partial_d q_{ab}\right)\,.\label{Gamma} 
\end{equation}
On the other hand, using Eq.~\eqref{eqALG} we obtain
\begin{eqnarray}
 q^{ab}&=&\frac{(1+\kappa\Lambda) g^{ab}-\kappa T^{ab}}{\sqrt{g}\sqrt{\det{\left((1+\kappa\Lambda) g^{ab}-\kappa T^{ab}\right)}}}\,,\label{quu}
\end{eqnarray}
which can be rewritten as
\begin{eqnarray}
 \kappa {\cal R}_{(ab)}&=&\sqrt{g}\sqrt{\det{\left[(1+\kappa\Lambda) g^{ab}-\kappa T^{ab}\right]}}\nn\\
&&\times\left[(1+\kappa\Lambda) g^{ab}-\kappa T^{ab}\right]^{-1}-g_{ab}\,.\label{eqg_exact}
\end{eqnarray}
Equation~\eqref{quu} determines $q_{ab}$ {\em algebraically} in terms of $g_{ab}$ and $T_{ab}$, whereas Eq.~\eqref{Gamma} determines $\Gamma_{ab}^c$ as the Levi-Civita connection of $q_{ab}$. Hence, one can use these equations to eliminate $\Gamma_{ab}^c$ from Eq.~\eqref{eqg_exact}. Then, the latter is the equation that has to be used to determine $g_{ab}$.

It is straightforward to see that, after eliminating $\Gamma_{ab}^c$, Eq.~(\ref{eqg_exact}) becomes a second-order partial differential equation in $g_{ab}$. However, it is equally straightforward to see that it also contains second derivatives of $T_{ab}$.
This is true in the full theory, but it becomes more explicit if we expand Eq.~\eqref{eqg_exact} at first order in $\kappa$. The expansion can be found easily by noting that
\begin{eqnarray}
 q^{ab}&=&g^{ab}-\kappa\tau^{ab}+{\cal O}(\kappa^2) \,,\nn
\end{eqnarray}
where  $\tau_{ab}\equiv T_{ab}-\frac{1}{2}g_{ab}T+\Lambda g_{ab}$. Using the expression above, we get a single equation for the metric $g_{ab}$ only:
\begin{eqnarray}
 R_{ab}&&=\Lambda g_{ab}+T_{ab}-\frac{1}{2}Tg_{ab}+\kappa\left[S_{ab}-\frac{1}{4}Sg_{ab}\right]\nn\\
&&+\frac{\kappa}{2}\left[\nabla_a\nabla_b \tau-2\nabla^c\nabla_{(a}\tau_{cb)}+\square\tau_{ab}\right]+{\cal O}(\kappa^2)\,. \label{eqg2}
\end{eqnarray}
where $S_{ab}={T^c}_a T_{c b}-\frac{1}{2}T T_{ab}$ and we have used the fact that $\tau_{ab}$ is symmetric. Note that now $R_{ab}$ is built solely from the Levi-Civita connection of $g_{ab}$ and $\nabla_a$ is the covariant derivative associated with $g_{ab}$. Therefore, $\Lambda$ inside $\tau_{ab}$ does not contribute to the second line of Eq.~\eqref{eqg2}. 

General relativity corresponds to $\kappa= 0$ and for $\kappa\neq 0$ Eq.~\eqref{eqg2} contains second derivatives of $T_{ab}$, {\em i.e.}~at least \emph{third} derivatives of the matter fields (unless we consider a fluid approximation of matter).  This is in contrast to Einstein's theory, where usually only first derivatives of the matter fields appear on the right hand side of Einstein's equations. 
This different structure is also evident in the Newtonian limit of the theory~\cite{Banados:2010ix}, which can be straightforwardly obtained from Eq.~\eqref{eqg2} but the results holds for any value of $\kappa$ and does not hinge on the small $\kappa$ expansion. The modified Poisson equation is $\nabla^2\Phi=\frac{\rho}{2}+\frac{\kappa}{4}\nabla^2\rho$, whose solution reads
%
\begin{equation}
\Phi=\Phi_{\rm N}+\frac{\kappa}{4}\rho\,,\label{Poisson_sol}
\end{equation}
where $\rho$ is the matter density and $\Phi_{\rm N}$ is the standard Newtonian potential.
Although the modified Newtonian regime has been studied in some detail and it provides interesting phenomenology~\cite{Pani:2012qb,Casanellas:2011kf}, Eq.~\eqref{Poisson_sol} shows that the gravitational potential $\Phi$ is algebraically related to~$\rho$.
This demonstrates that gravity is noncommulative: unlike in Einstein's theory, the metric in EiBI gravity is not just an integral over the sources but it receives an algebraic contribution from the matter fields and their derivatives. 
Any matter configuration which is discontinuous or just not smooth enough will produce discontinuities in the metric and singularities in the curvature invariants (which depend on the second derivatives of~$\Phi$), leading to unacceptable phenomenology.
Clearly, this behaviour persist in the post-Newtonian limit (cf. Ref.~\cite{Sotiriou:2008rp} for a discussion in Palatini $f({\cal R})$ gravity).
These problems have been overlooked in the literature of EiBI theory. 

As mentioned earlier, the same pathologies arise in Palatini $f({\cal R})$ gravity~\cite{Sotiriou:2008rp} and in theories with matter coupled to the Ricci scalar~\cite{Sotiriou:2008dh}.
Indeed, the structure of Eq.~\eqref{eqg2} is the same as in Palatini $f({\cal R})$ theory (cf. Eq.~(28) in Ref.~\cite{Sotiriou:2008rp}). A qualitative difference is that our Eq.~\eqref{eqg2} contains derivatives of the full stress-energy tensor, whereas the field equations in Palatini $f({\cal R})$ gravity only contain derivatives of the trace $T=g^{ab}T_{ab}$. Thus, the problems we are discussing are already manifest when EiBI gravity is simply coupled to a Maxwell field, whose stress-energy tensor is traceless, $T\equiv0$.

Let us now discuss some problems in constructing perfect-fluid equilibrium structures in EiBI gravity, which are related to its peculiar differential structure. We shall build on previous works which discuss similar pathologies in other theories~\cite{Sotiriou:2008rp,Sotiriou:2008dh}.
Static and spherically symmetric perfect-fluid stars were discussed in detail in Refs.~\cite{Pani:2011mg,Pani:2012qb}. 
Without loss of generality, a convenient ansatz for the metrics reads
\bea
q_{ab}dx^a dx^b &=& -p(r) dt^2 + h(r) dr^2 + r^2 d\Omega^2,\label{ansatzq}\\
g_{ab}dx^a dx^b &=& -F(r) dt^2 + B(r) dr^2 + A(r)r^2 d\Omega^2\,,\label{ansatzg}
\eea
where we have used the gauge freedom to fix the function in front of the spherical part of the auxiliary $q_{ab}$ metric.
We consider perfect-fluid stars whose stress-energy tensor reads $T_{ab}=\left[\rho+P\right]u_a u_b+g_{ab}P$,
where $u^a=(1/\sqrt{F},0,0,0)$ and $\rho(r)$ and $P(r)$ denote the energy density and the pressure respectively.

Notice that the field equations~\eqref{eqALG} are simply algebraic relations between $q_{ab}$ and $g_{ab}$. Inserting the ans\"atze above into Eqs.~\eqref{eqALG}, one can solve for the coefficients of $g_{ab}$ in terms of $q_{ab}$ and $T_{ab}$~\cite{Pani:2012qb}. Then, the dynamical equations~\eqref{eqg_exact} do not explicitly contain second derivatives of the matter fields and can be solved for $p$, $h$ and $P$, assuming an equation of state of the form $P=P(\rho)$.
Interestingly, in this formulation the field equations are equivalent to the standard Tolman---Oppenheimer---Volkoff equations for the metric $q_{ab}$, with an effective equation of state~\cite{Delsate:2012ky}.
Since matter is covariantly coupled to the $g_{ab}$ metric, the standard conservation of the stress-energy tensor follows, $\nabla_a T^{ab}=0$. Finally, the interior solution is matched to the (unique) exterior Schwarzschild metric through appropriate junction conditions at the stellar surface~\cite{Pani:2011mg,Pani:2012qb}.
However, in this formulation the physical $g_{ab}$ metric is not a dynamical quantity and a successful numerical integration does not necessarily mean that geometric invariants, which involve \emph{derivatives} of $g_{ab}$, are regular. Indeed, we show here that the Ricci curvature of the physical $g_{ab}$ metric is divergent at the surface. This singularity has been missed in previous literature on EiBI gravity. 

For simplicity, we consider the asymptotically flat case, $\Lambda=0$. Using Eqs.~\eqref{ansatzq}--\eqref{ansatzg}, the Ricci curvature reads
\begin{eqnarray}
 R_g&=&\left[{2 r^2 A^2 B^2 F^2}\right]^{-1}\left\{r^2 B F^2 A'^2+2 A F \left[2 B^2 F\right.\right.\nn\\
&&\left.\left.+r^2 F A' B'-r B \left(A' \left(6 F+r F'\right)+2 r F A''\right)\right]\right.\nn\\
&&\left.+A^2 \left[r F B' \left(4 F+r F'\right)+B \left(-4 F^2+r^2 F'^2\right.\right.\right.\nn\\
&&\left.\left.\left.-2 r F \left(2 F'+r F''\right)\right)\right]\right\}\,,\label{ricciscal}
\end{eqnarray}
which involves first and second derivatives of the $g_{ab}$ metric coefficients $F$, $B$ and $A$. By using the algebraic relations, we can write $R_g$ only in terms of the $q_{ab}$-metric coefficients $p$ and $h$ and of the matter fields $P$ and $\rho$. The final expression can be schematically written as
\begin{equation}
 R_g=R_g(p,p',p'',h,h',P,P',P'')\,,\label{ricciscal_impl}
\end{equation}
where we have used the equation of state to eliminate $\rho$ and its derivatives.
In general relativity the Ricci curvature simply reads $R_g=-T=\rho-3P$, {\em i.e.}~no derivatives of the matter fields appear. This has profound implications. For example, if the function $P(r)$ is continuous but not differentiable at the stellar surface, then $P'$ would be discontinuous at the radius and $P''$ would introduce an unacceptable Dirac delta contribution to the curvature. However, the differentiability of $P(r)$ at the surface is hard to judge before having solved the field equations.

In the specific case of polytropic equations of state however, where $P=K\rho_0^\Gamma$, with $\rho_0$ being the rest-mass density and $K$ and $\Gamma$ constants, one can determine the behaviour of $R_g$ at the surface without actually having to solve the equations fully. The energy density can then be written explicitly as $\rho(P)=\left[{P}/{K}\right]^{1/\Gamma}+{P}/(\Gamma-1)$.  We can use the field equations~\cite{Pani:2012qb} to eliminate the derivatives in Eq.~\eqref{ricciscal_impl}.  Evaluating $R_g$ at the stellar surface, {\em i.e.} as $r\to R_S$ and $P\to0$, for any $\kappa\neq0$  we get 
\begin{equation}
 R_g(P\to0)\sim \left\{
 \begin{array}{l}
  \gamma_\Gamma\,, 						\hspace{2.1cm} 	0<\Gamma\leq3/2 \\
  \gamma_\Gamma P^{-2+{3}/{\Gamma }}\,,  		\hspace{0.4cm}	3/2<\Gamma<2 \\
  \gamma_\Gamma P^{-1/\Gamma}\,,			\hspace{1.75cm}	\Gamma\geq2\\
 \end{array}\right.
\,.\label{Rsing}
\end{equation}
In the equation above, $\gamma_\Gamma=0$ if $0<\Gamma<3/2$, whereas
\begin{equation}
 \gamma_\Gamma=\left\{
 \begin{array}{l}
  \frac{\kappa  (2-\Gamma)}{2\eta K^{3/\Gamma}  \Gamma ^2 }\,, 	\hspace{1.35cm} 	3/2\leq\Gamma<2 \vspace{.15cm}\\
  \frac{-8 \kappa^2 }{[8+\kappa/ K]^3\eta K^{5/2} }\,,  	\hspace{1.6cm}	\Gamma=2 \vspace{.15cm}\\\
  \frac{8(1-\Gamma) K^{1/\Gamma}}{\kappa \eta }\,,		\hspace{1.95cm}	\Gamma>2\\
 \end{array}\right.
\,,
\end{equation}
with $\eta=R_S^3(R_S-2M)/M^2$ and $M$ is the total mass defined by $h(R_S)=[1-2M/R_S]^{-1}$~\cite{Pani:2012qb,Sham:2012qi}. 
Therefore, for \emph{any} $\Gamma>3/2$ the scalar curvature diverges at the surface. 
The diverging terms originate from the derivatives of the matter fields in Eq.~\eqref{ricciscal_impl} and more specifically from  the terms $\sim\kappa\rho^2\rho_{PP}$, $\kappa\rho \rho_P$ and $\rho_{PP}/(\kappa\rho_P^2)$, where the subscripts denote partial derivatives with respect to $P$.

A similar result, {\em i.e.} the divergence of the Ricci scalar at the surface of polytropic matter configurations, was obtained in Palatini $f({\cal R})$ gravity and the consequences are discussed in detail in Refs.~\cite{Barausse:2007pn,Barausse:2007ys,Barausse:2008nm}. In that case, the divergence occurs only for $3/2<\Gamma<2$. In EiBI gravity one has $\gamma_\Gamma\sim1/\kappa$ for $\Gamma>2$, so that this singular solution does not appear in a small $\kappa$ expansion in which EiBI gravity resembles Palatini $f({\cal R},{\cal R}_{ab})$.

What we have established is that EiBI gravity does not admit {\em any} regular solution for polytropic spheres with $\Gamma>3/2$, even for arbitrarily small values of $\kappa$. 
At least two physical matter configurations are exactly described by a polytropic equation of state with $\Gamma>3/2$: a degenerate gas of nonrelativistic electrons and a monoatomic isentropic gas, both having $\Gamma=5/3$.
These perfectly reasonable systems, which can even be described within Newtonian theory, have no description in EiBI theory. This renders the theory at best incomplete. 

It is worth stressing that our analysis relies only on the field equations and on the form of the equation of state \emph{close} to the stellar surface. Any matter configuration whose behaviour resembles a polytrope (as an effective description) with adequate accuracy in the immediate vicinity of the surface will be singular, regardless of any complicated microphysics describing the stellar interior. There are many known examples of stars that satisfy this property. For instance the atmosphere of a white dwarf is well approximated by a polytrope with $\Gamma=10/3$ (see e.g. Refs.~\cite{ZeldovichNovikov,Shapiro:1983du}).

At this stage one might claim that the polytropic, perfect-fluid description will break down at very small densities, and this might be a potential way out. Nonetheless, this would imply that no solutions for white dwarfs (and for many other systems) are allowed in EiBI gravity without precise knowledge of the microphysics of the matter near the surface. Even after introducing some microphysical description, there are no guaranties that the solution would be regular. 
Indeed abandoning the fluid approximation would just increase the differential order of the field equations in the matter sector, making the curvature even more sensitive to sharp variations in the matter fields.

For the sake of the argument though, let us suppose that a polytropic equation of state provides a reliable description close to the surface down to, say $\rho\sim 10^{-n}\text{kg/m}^3$. 
Strong deviations from general relativity are expected when ${R_g^\text{EiBI}}\gg{R_g^\text{GR}}=\rho-3P$ at some radius very near the surface. For example, surface singularities would give rise to divergent tidal forces, which can be orders of magnitude larger than in Einstein's theory~\cite{Barausse:2007ys}.
Let us then require that ${R_g^\text{EiBI}}\lesssim {R_g^\text{GR}}$. This yields
\begin{equation}
 \kappa\gtrsim 4\times10^{24+2n}\text{m}^5\text{kg}^{-1}\text{s}^{-2}\,.\label{bound}
\end{equation}
where we have assumed $M\sim1.4M_\odot$, $R_S\sim10^{-2}R_\odot$ and $\Gamma=10/3$. 
For $n=10$, the absence of strong near-surface curvature effects would imply $\kappa\gtrsim 4\times10^{44}\text{m}^5\text{kg}^{-1}\text{s}^{-2}$. This bound, though admittedly simplistic, is nonetheless about 40 orders of magnitude larger than other current constraints~\cite{Pani:2011mg,Casanellas:2011kf,Avelino:2012ge,DeFelice:2012hq,Avelino:2012qe}. Additionally, Eq.~\eqref{bound} is a \emph{lower} bound whereas all other constraints (including those one could derive by applying similar arguments to matter configurations well described by polytropes with $3/2<\Gamma<2$) are upper bounds that are totally incompatible with Eq.~\eqref{bound}.

We emphasize that the surface singularities found above are not a prerogative of some polytropic fluid description of matter near the surface. Polytropes just allow for a convenient analytical treatment of the problem. The key issue is that higher-order derivatives of matter fields, which appear in the EiBI field equations as a result of integrating out nondynamical degrees of freedom, make the geometry unacceptably sensitive to sharp variations in the matter configuration. Surface singularities in polytropic spheres are just one possible manifestation of this sensitivity,
but similar phenomena will be present in real world systems, where sharp density variations are common. This shortcoming, the rest of the pathologies related to the presence of nondynamical fields and a thorough discussion on the limitations of the polytropic fluid approximation can 
been found in Ref.~\cite{Barausse:2008nm} for Palatini  $f({\cal R})$ gravity (see also Refs.~\cite{Kainulainen:2007bt,Olmo:2008pv}). We will avoid repeating this discussion here for EiBI gravity, as it would be nearly identical. 

In summary, we have shown that EiBI gravity  is plagued by serious pathologies, whose root is the fact that the theory contains an auxiliary connection. The latter can be eliminated in order to obtain second-order dynamical equations for the metric only. However, the differential structure of these equations is profoundly different from general relativity, as higher derivatives of the matter fields appear. This makes gravity noncommulative and spacetime geometry particularly sensitive to sharp changes in the matter configuration. In particular, a discontinuity in the matter density or even only in its derivatives is enough to produce curvature singularities,  leading to unacceptable phenomenology.
This fact has been missed in the recent literature on EiBI gravity but it has profound consequences for the viability of the theory, similarly to the case of Palatini $f({\cal R})$ gravities~\cite{Barausse:2007pn,Barausse:2007ys,Barausse:2008nm} and 
to theories 
where matter is coupled to the Ricci scalar~\cite{Sotiriou:2008dh}. 

These problems appear to be a generic prerogative of gravitational theories which do not propagate any degree of freedom other than the massless spin-2 field, but instead contain auxiliary fields that are algebraically related to the metric and to matter. Although these theories are equivalent to general relativity in vacuum, once the auxiliary fields are eliminated using some algebraic relation, the dynamical gravitational equations contain higher derivatives of the matter fields. 
This is the case discussed here, where the auxiliary field is the connection or, equivalently, the $q_{ab}$ metric. 

In principle, this pathology could  be alleviated by introducing higher order derivatives of the metric as well (this is for example the case in which matter is coupled to some \emph{nonlinear} function of the Ricci scalar~\cite{Bertolami:2007gv}). However, this usually produces different problems, like unitarity loss, instabilities, etc.

Finally, in this work we followed the original proposal, assuming the independent connection to be symmetric, the matter action to be independent from it and that only ${\cal R}_{(ab)}$ appears in the action. Relaxing these conditions would lead to the most general, metric-affine~\cite{Sotiriou:2006qn} version of EiBI gravity, which presumably has a richer phenomenology. 
Understanding whether this more general theory may evade the pathologies discussed here would be an interesting extension of the present work. 

{\em Acknowledgments}:
  We wish to thank T.~Delsate, J.~Steinhoff, V.~Vitagliano and especially E.~Barausse for useful discussions.
  PP is supported by FCT--Portugal through projects PTDC/FIS/098025/2008, PTDC/FIS/098032/2008, CERN/FP/123593/2011 and by
  the European Community through the Intra-European Marie Curie contract
  aStronGR-2011-298297. TPS acknowledges partial financial
support provided under the Marie Curie Career Integration
Grant 	LIMITSOFGR-2011-TPS, the ``Young SISSA Scientists¢ Research
Project'' scheme 2011-2012, and the European Union's FP7 ERC grant agreement no. 306425.

\bibliography{Eddington}

\begin{thebibliography}{31}%
\makeatletter
\providecommand \@ifxundefined [1]{%
 \@ifx{#1\undefined}
}%
\providecommand \@ifnum [1]{%
 \ifnum #1\expandafter \@firstoftwo
 \else \expandafter \@secondoftwo
 \fi
}%
\providecommand \@ifx [1]{%
 \ifx #1\expandafter \@firstoftwo
 \else \expandafter \@secondoftwo
 \fi
}%
\providecommand \natexlab [1]{#1}%
\providecommand \enquote  [1]{``#1''}%
\providecommand \bibnamefont  [1]{#1}%
\providecommand \bibfnamefont [1]{#1}%
\providecommand \citenamefont [1]{#1}%
\providecommand \href@noop [0]{\@secondoftwo}%
\providecommand \href [0]{\begingroup \@sanitize@url \@href}%
\providecommand \@href[1]{\@@startlink{#1}\@@href}%
\providecommand \@@href[1]{\endgroup#1\@@endlink}%
\providecommand \@sanitize@url [0]{\catcode `\\12\catcode `\$12\catcode
  `\&12\catcode `\#12\catcode `\^12\catcode `\_12\catcode `\%12\relax}%
\providecommand \@@startlink[1]{}%
\providecommand \@@endlink[0]{}%
\providecommand \url  [0]{\begingroup\@sanitize@url \@url }%
\providecommand \@url [1]{\endgroup\@href {#1}{\urlprefix }}%
\providecommand \urlprefix  [0]{URL }%
\providecommand \Eprint [0]{\href }%
\providecommand \doibase [0]{http://dx.doi.org/}%
\providecommand \selectlanguage [0]{\@gobble}%
\providecommand \bibinfo  [0]{\@secondoftwo}%
\providecommand \bibfield  [0]{\@secondoftwo}%
\providecommand \translation [1]{[#1]}%
\providecommand \BibitemOpen [0]{}%
\providecommand \bibitemStop [0]{}%
\providecommand \bibitemNoStop [0]{.\EOS\space}%
\providecommand \EOS [0]{\spacefactor3000\relax}%
\providecommand \BibitemShut  [1]{\csname bibitem#1\endcsname}%
\let\auto@bib@innerbib\@empty
\bibitem [{\citenamefont {Clifton}\ \emph {et~al.}(2012)\citenamefont
  {Clifton}, \citenamefont {Ferreira}, \citenamefont {Padilla},\ and\
  \citenamefont {Skordis}}]{Clifton:2011jh}%
  \BibitemOpen
  \bibfield  {author} {\bibinfo {author} {\bibfnamefont {T.}~\bibnamefont
  {Clifton}}, \bibinfo {author} {\bibfnamefont {P.~G.}\ \bibnamefont
  {Ferreira}}, \bibinfo {author} {\bibfnamefont {A.}~\bibnamefont {Padilla}}, \
  and\ \bibinfo {author} {\bibfnamefont {C.}~\bibnamefont {Skordis}},\
  }\href@noop {} {\bibfield  {journal} {\bibinfo  {journal} {Phys.Rept.}\
  }\textbf {\bibinfo {volume} {513}},\ \bibinfo {pages} {1} (\bibinfo {year}
  {2012})},\ \Eprint {http://arxiv.org/abs/1106.2476} {arXiv:1106.2476
  [astro-ph.CO]} \BibitemShut {NoStop}%
\bibitem [{\citenamefont {Banados}\ and\ \citenamefont
  {Ferreira}(2010)}]{Banados:2010ix}%
  \BibitemOpen
  \bibfield  {author} {\bibinfo {author} {\bibfnamefont {M.}~\bibnamefont
  {Banados}}\ and\ \bibinfo {author} {\bibfnamefont {P.~G.}\ \bibnamefont
  {Ferreira}},\ }\href {\doibase 10.1103/PhysRevLett.105.011101} {\bibfield
  {journal} {\bibinfo  {journal} {Phys.Rev.Lett.}\ }\textbf {\bibinfo {volume}
  {105}},\ \bibinfo {pages} {011101} (\bibinfo {year} {2010})}\BibitemShut
  {NoStop}%
\bibitem [{\citenamefont {Pani}\ \emph {et~al.}(2011)\citenamefont {Pani},
  \citenamefont {Cardoso},\ and\ \citenamefont {Delsate}}]{Pani:2011mg}%
  \BibitemOpen
  \bibfield  {author} {\bibinfo {author} {\bibfnamefont {P.}~\bibnamefont
  {Pani}}, \bibinfo {author} {\bibfnamefont {V.}~\bibnamefont {Cardoso}}, \
  and\ \bibinfo {author} {\bibfnamefont {T.}~\bibnamefont {Delsate}},\ }\href
  {\doibase 10.1103/PhysRevLett.107.031101} {\bibfield  {journal} {\bibinfo
  {journal} {Phys.Rev.Lett.}\ }\textbf {\bibinfo {volume} {107}},\ \bibinfo
  {pages} {031101} (\bibinfo {year} {2011})},\ \Eprint
  {http://arxiv.org/abs/1106.3569} {arXiv:1106.3569 [gr-qc]} \BibitemShut
  {NoStop}%
\bibitem [{\citenamefont {Casanellas}\ \emph {et~al.}(2012)\citenamefont
  {Casanellas}, \citenamefont {Pani}, \citenamefont {Lopes},\ and\
  \citenamefont {Cardoso}}]{Casanellas:2011kf}%
  \BibitemOpen
  \bibfield  {author} {\bibinfo {author} {\bibfnamefont {J.}~\bibnamefont
  {Casanellas}}, \bibinfo {author} {\bibfnamefont {P.}~\bibnamefont {Pani}},
  \bibinfo {author} {\bibfnamefont {I.}~\bibnamefont {Lopes}}, \ and\ \bibinfo
  {author} {\bibfnamefont {V.}~\bibnamefont {Cardoso}},\ }\href {\doibase
  10.1088/0004-637X/745/1/15} {\bibfield  {journal} {\bibinfo  {journal}
  {Astrophys.J.}\ }\textbf {\bibinfo {volume} {745}},\ \bibinfo {pages} {15}
  (\bibinfo {year} {2012})},\ \Eprint {http://arxiv.org/abs/1109.0249}
  {arXiv:1109.0249 [astro-ph.SR]} \BibitemShut {NoStop}%
\bibitem [{\citenamefont {Avelino}(2012{\natexlab{a}})}]{Avelino:2012ge}%
  \BibitemOpen
  \bibfield  {author} {\bibinfo {author} {\bibfnamefont {P.}~\bibnamefont
  {Avelino}},\ }\href {\doibase 10.1103/PhysRevD.85.104053} {\bibfield
  {journal} {\bibinfo  {journal} {Phys.Rev.}\ }\textbf {\bibinfo {volume}
  {D85}},\ \bibinfo {pages} {104053} (\bibinfo {year} {2012}{\natexlab{a}})},\
  \Eprint {http://arxiv.org/abs/1201.2544} {arXiv:1201.2544 [astro-ph.CO]}
  \BibitemShut {NoStop}%
\bibitem [{\citenamefont {Pani}\ \emph {et~al.}(2012)\citenamefont {Pani},
  \citenamefont {Delsate},\ and\ \citenamefont {Cardoso}}]{Pani:2012qb}%
  \BibitemOpen
  \bibfield  {author} {\bibinfo {author} {\bibfnamefont {P.}~\bibnamefont
  {Pani}}, \bibinfo {author} {\bibfnamefont {T.}~\bibnamefont {Delsate}}, \
  and\ \bibinfo {author} {\bibfnamefont {V.}~\bibnamefont {Cardoso}},\ }\href
  {\doibase 10.1103/PhysRevD.85.084020} {\bibfield  {journal} {\bibinfo
  {journal} {Phys. Rev. D}\ }\textbf {\bibinfo {volume} {85}},\ \bibinfo
  {pages} {084020} (\bibinfo {year} {2012})}\BibitemShut {NoStop}%
\bibitem [{\citenamefont {Delsate}\ and\ \citenamefont
  {Steinhoff}(2012)}]{Delsate:2012ky}%
  \BibitemOpen
  \bibfield  {author} {\bibinfo {author} {\bibfnamefont {T.}~\bibnamefont
  {Delsate}}\ and\ \bibinfo {author} {\bibfnamefont {J.}~\bibnamefont
  {Steinhoff}},\ }\href {\doibase 10.1103/PhysRevLett.109.021101} {\bibfield
  {journal} {\bibinfo  {journal} {Phys.Rev.Lett.}\ }\textbf {\bibinfo {volume}
  {109}},\ \bibinfo {pages} {021101} (\bibinfo {year} {2012})},\ \Eprint
  {http://arxiv.org/abs/1201.4989} {arXiv:1201.4989 [gr-qc]} \BibitemShut
  {NoStop}%
\bibitem [{\citenamefont {Liu}\ \emph {et~al.}(2012)\citenamefont {Liu},
  \citenamefont {Yang}, \citenamefont {Guo},\ and\ \citenamefont
  {Zhong}}]{Liu:2012rc}%
  \BibitemOpen
  \bibfield  {author} {\bibinfo {author} {\bibfnamefont {Y.-X.}\ \bibnamefont
  {Liu}}, \bibinfo {author} {\bibfnamefont {K.}~\bibnamefont {Yang}}, \bibinfo
  {author} {\bibfnamefont {H.}~\bibnamefont {Guo}}, \ and\ \bibinfo {author}
  {\bibfnamefont {Y.}~\bibnamefont {Zhong}},\ }\href {\doibase
  10.1103/PhysRevD.85.124053} {\bibfield  {journal} {\bibinfo  {journal}
  {Phys.Rev.}\ }\textbf {\bibinfo {volume} {D85}},\ \bibinfo {pages} {124053}
  (\bibinfo {year} {2012})},\ \Eprint {http://arxiv.org/abs/1203.2349}
  {arXiv:1203.2349 [hep-th]} \BibitemShut {NoStop}%
\bibitem [{\citenamefont {Escamilla-Rivera}\ \emph {et~al.}(2012)\citenamefont
  {Escamilla-Rivera}, \citenamefont {Banados},\ and\ \citenamefont
  {Ferreira}}]{EscamillaRivera:2012vz}%
  \BibitemOpen
  \bibfield  {author} {\bibinfo {author} {\bibfnamefont {C.}~\bibnamefont
  {Escamilla-Rivera}}, \bibinfo {author} {\bibfnamefont {M.}~\bibnamefont
  {Banados}}, \ and\ \bibinfo {author} {\bibfnamefont {P.~G.}\ \bibnamefont
  {Ferreira}},\ }\href@noop {} {\bibfield  {journal} {\bibinfo  {journal}
  {Phys.Rev.}\ }\textbf {\bibinfo {volume} {D85}},\ \bibinfo {pages} {087302}
  (\bibinfo {year} {2012})},\ \Eprint {http://arxiv.org/abs/1204.1691}
  {arXiv:1204.1691 [gr-qc]} \BibitemShut {NoStop}%
\bibitem [{\citenamefont {Avelino}\ and\ \citenamefont
  {Ferreira}(2012)}]{Avelino:2012ue}%
  \BibitemOpen
  \bibfield  {author} {\bibinfo {author} {\bibfnamefont {P.}~\bibnamefont
  {Avelino}}\ and\ \bibinfo {author} {\bibfnamefont {R.}~\bibnamefont
  {Ferreira}},\ }\href {\doibase 10.1103/PhysRevD.86.041501} {\bibfield
  {journal} {\bibinfo  {journal} {Phys.Rev.}\ }\textbf {\bibinfo {volume}
  {D86}},\ \bibinfo {pages} {041501} (\bibinfo {year} {2012})},\ \Eprint
  {http://arxiv.org/abs/1205.6676} {arXiv:1205.6676 [astro-ph.CO]} \BibitemShut
  {NoStop}%
\bibitem [{\citenamefont {Avelino}(2012{\natexlab{b}})}]{Avelino:2012qe}%
  \BibitemOpen
  \bibfield  {author} {\bibinfo {author} {\bibfnamefont {P.}~\bibnamefont
  {Avelino}},\ }\href@noop {} {\  (\bibinfo {year} {2012}{\natexlab{b}})},\
  \Eprint {http://arxiv.org/abs/1207.4730} {arXiv:1207.4730 [astro-ph.CO]}
  \BibitemShut {NoStop}%
\bibitem [{\citenamefont {Sham}\ \emph {et~al.}(2012)\citenamefont {Sham},
  \citenamefont {Lin},\ and\ \citenamefont {Leung}}]{Sham:2012qi}%
  \BibitemOpen
  \bibfield  {author} {\bibinfo {author} {\bibfnamefont {Y.-H.}\ \bibnamefont
  {Sham}}, \bibinfo {author} {\bibfnamefont {L.-M.}\ \bibnamefont {Lin}}, \
  and\ \bibinfo {author} {\bibfnamefont {P.}~\bibnamefont {Leung}},\
  }\href@noop {} {\  (\bibinfo {year} {2012})},\ \Eprint
  {http://arxiv.org/abs/1208.1314} {arXiv:1208.1314 [gr-qc]} \BibitemShut
  {NoStop}%
\bibitem [{\citenamefont {Cho}\ \emph {et~al.}(2012)\citenamefont {Cho},
  \citenamefont {Kim},\ and\ \citenamefont {Moon}}]{Cho:2012vg}%
  \BibitemOpen
  \bibfield  {author} {\bibinfo {author} {\bibfnamefont {I.}~\bibnamefont
  {Cho}}, \bibinfo {author} {\bibfnamefont {H.-C.}\ \bibnamefont {Kim}}, \ and\
  \bibinfo {author} {\bibfnamefont {T.}~\bibnamefont {Moon}},\ }\href@noop {}
  {\  (\bibinfo {year} {2012})},\ \Eprint {http://arxiv.org/abs/1208.2146}
  {arXiv:1208.2146 [gr-qc]} \BibitemShut {NoStop}%
\bibitem [{\citenamefont {De~Felice}\ \emph {et~al.}(2012)\citenamefont
  {De~Felice}, \citenamefont {Gumjudpai},\ and\ \citenamefont
  {Jhingan}}]{DeFelice:2012hq}%
  \BibitemOpen
  \bibfield  {author} {\bibinfo {author} {\bibfnamefont {A.}~\bibnamefont
  {De~Felice}}, \bibinfo {author} {\bibfnamefont {B.}~\bibnamefont
  {Gumjudpai}}, \ and\ \bibinfo {author} {\bibfnamefont {S.}~\bibnamefont
  {Jhingan}},\ }\href {\doibase 10.1103/PhysRevD.86.043525} {\bibfield
  {journal} {\bibinfo  {journal} {Phys.Rev.}\ }\textbf {\bibinfo {volume}
  {D86}},\ \bibinfo {pages} {043525} (\bibinfo {year} {2012})},\ \Eprint
  {http://arxiv.org/abs/1205.1168} {arXiv:1205.1168 [gr-qc]} \BibitemShut
  {NoStop}%
\bibitem [{\citenamefont {Sotiriou}\ and\ \citenamefont
  {Faraoni}(2010)}]{Sotiriou:2008rp}%
  \BibitemOpen
  \bibfield  {author} {\bibinfo {author} {\bibfnamefont {T.~P.}\ \bibnamefont
  {Sotiriou}}\ and\ \bibinfo {author} {\bibfnamefont {V.}~\bibnamefont
  {Faraoni}},\ }\href {\doibase 10.1103/RevModPhys.82.451} {\bibfield
  {journal} {\bibinfo  {journal} {Rev.Mod.Phys.}\ }\textbf {\bibinfo {volume}
  {82}},\ \bibinfo {pages} {451} (\bibinfo {year} {2010})},\ \Eprint
  {http://arxiv.org/abs/0805.1726} {arXiv:0805.1726 [gr-qc]} \BibitemShut
  {NoStop}%
\bibitem [{\citenamefont {Flanagan}(2004)}]{Flanagan:2003rb}%
  \BibitemOpen
  \bibfield  {author} {\bibinfo {author} {\bibfnamefont {E.~E.}\ \bibnamefont
  {Flanagan}},\ }\href {\doibase 10.1103/PhysRevLett.92.071101} {\bibfield
  {journal} {\bibinfo  {journal} {Phys.Rev.Lett.}\ }\textbf {\bibinfo {volume}
  {92}},\ \bibinfo {pages} {071101} (\bibinfo {year} {2004})},\ \Eprint
  {http://arxiv.org/abs/astro-ph/0308111} {arXiv:astro-ph/0308111 [astro-ph]}
  \BibitemShut {NoStop}%
\bibitem [{\citenamefont {Barausse}\ \emph
  {et~al.}(2008{\natexlab{a}})\citenamefont {Barausse}, \citenamefont
  {Sotiriou},\ and\ \citenamefont {Miller}}]{Barausse:2007pn}%
  \BibitemOpen
  \bibfield  {author} {\bibinfo {author} {\bibfnamefont {E.}~\bibnamefont
  {Barausse}}, \bibinfo {author} {\bibfnamefont {T.~P.}\ \bibnamefont
  {Sotiriou}}, \ and\ \bibinfo {author} {\bibfnamefont {J.~C.}\ \bibnamefont
  {Miller}},\ }\href {\doibase 10.1088/0264-9381/25/6/062001} {\bibfield
  {journal} {\bibinfo  {journal} {Class.Quant.Grav.}\ }\textbf {\bibinfo
  {volume} {25}},\ \bibinfo {pages} {062001} (\bibinfo {year}
  {2008}{\natexlab{a}})},\ \Eprint {http://arxiv.org/abs/gr-qc/0703132}
  {arXiv:gr-qc/0703132 [GR-QC]} \BibitemShut {NoStop}%
\bibitem [{\citenamefont {Barausse}\ \emph
  {et~al.}(2008{\natexlab{b}})\citenamefont {Barausse}, \citenamefont
  {Sotiriou},\ and\ \citenamefont {Miller}}]{Barausse:2007ys}%
  \BibitemOpen
  \bibfield  {author} {\bibinfo {author} {\bibfnamefont {E.}~\bibnamefont
  {Barausse}}, \bibinfo {author} {\bibfnamefont {T.}~\bibnamefont {Sotiriou}},
  \ and\ \bibinfo {author} {\bibfnamefont {J.}~\bibnamefont {Miller}},\ }\href
  {\doibase 10.1088/0264-9381/25/10/105008} {\bibfield  {journal} {\bibinfo
  {journal} {Class.Quant.Grav.}\ }\textbf {\bibinfo {volume} {25}},\ \bibinfo
  {pages} {105008} (\bibinfo {year} {2008}{\natexlab{b}})},\ \Eprint
  {http://arxiv.org/abs/0712.1141} {arXiv:0712.1141 [gr-qc]} \BibitemShut
  {NoStop}%
\bibitem [{\citenamefont {Barausse}\ \emph
  {et~al.}(2008{\natexlab{c}})\citenamefont {Barausse}, \citenamefont
  {Sotiriou},\ and\ \citenamefont {Miller}}]{Barausse:2008nm}%
  \BibitemOpen
  \bibfield  {author} {\bibinfo {author} {\bibfnamefont {E.}~\bibnamefont
  {Barausse}}, \bibinfo {author} {\bibfnamefont {T.~P.}\ \bibnamefont
  {Sotiriou}}, \ and\ \bibinfo {author} {\bibfnamefont {J.~C.}\ \bibnamefont
  {Miller}},\ }\href {\doibase 10.1051/eas:0830023} {\bibfield  {journal}
  {\bibinfo  {journal} {EAS Publ.Ser.}\ }\textbf {\bibinfo {volume} {30}},\
  \bibinfo {pages} {189} (\bibinfo {year} {2008}{\natexlab{c}})},\ \Eprint
  {http://arxiv.org/abs/0801.4852} {arXiv:0801.4852 [gr-qc]} \BibitemShut
  {NoStop}%
\bibitem [{\citenamefont {Sotiriou}(2008)}]{Sotiriou:2008dh}%
  \BibitemOpen
  \bibfield  {author} {\bibinfo {author} {\bibfnamefont {T.~P.}\ \bibnamefont
  {Sotiriou}},\ }\href {\doibase 10.1016/j.physletb.2008.05.050} {\bibfield
  {journal} {\bibinfo  {journal} {Phys.Lett.}\ }\textbf {\bibinfo {volume}
  {B664}},\ \bibinfo {pages} {225} (\bibinfo {year} {2008})},\ \Eprint
  {http://arxiv.org/abs/0805.1160} {arXiv:0805.1160 [gr-qc]} \BibitemShut
  {NoStop}%
\bibitem [{\citenamefont {Olmo}(2010)}]{Olmo:2009xs}%
  \BibitemOpen
  \bibfield  {author} {\bibinfo {author} {\bibfnamefont {G.~J.}\ \bibnamefont
  {Olmo}},\ }\href {\doibase 10.1063/1.3462606} {\bibfield  {journal} {\bibinfo
   {journal} {AIP Conf.Proc.}\ }\textbf {\bibinfo {volume} {1241}},\ \bibinfo
  {pages} {1100} (\bibinfo {year} {2010})},\ \Eprint
  {http://arxiv.org/abs/0910.3734} {arXiv:0910.3734 [gr-qc]} \BibitemShut
  {NoStop}%
\bibitem [{\citenamefont {Deser}\ and\ \citenamefont
  {Gibbons}(1998)}]{Deser:1998rj}%
  \BibitemOpen
  \bibfield  {author} {\bibinfo {author} {\bibfnamefont {S.}~\bibnamefont
  {Deser}}\ and\ \bibinfo {author} {\bibfnamefont {G.}~\bibnamefont
  {Gibbons}},\ }\href {\doibase 10.1088/0264-9381/15/5/001} {\bibfield
  {journal} {\bibinfo  {journal} {Class.Quant.Grav.}\ }\textbf {\bibinfo
  {volume} {15}},\ \bibinfo {pages} {L35} (\bibinfo {year} {1998})},\ \Eprint
  {http://arxiv.org/abs/hep-th/9803049} {arXiv:hep-th/9803049 [hep-th]}
  \BibitemShut {NoStop}%
\bibitem [{\citenamefont {Vollick}(2004)}]{Vollick:2003qp}%
  \BibitemOpen
  \bibfield  {author} {\bibinfo {author} {\bibfnamefont {D.~N.}\ \bibnamefont
  {Vollick}},\ }\href {\doibase 10.1103/PhysRevD.69.064030} {\bibfield
  {journal} {\bibinfo  {journal} {Phys.Rev.}\ }\textbf {\bibinfo {volume}
  {D69}},\ \bibinfo {pages} {064030} (\bibinfo {year} {2004})},\ \Eprint
  {http://arxiv.org/abs/gr-qc/0309101} {arXiv:gr-qc/0309101 [gr-qc]}
  \BibitemShut {NoStop}%
\bibitem [{\citenamefont {Vitagliano}\ \emph {et~al.}(2010)\citenamefont
  {Vitagliano}, \citenamefont {Sotiriou},\ and\ \citenamefont
  {Liberati}}]{Vitagliano:2010pq}%
  \BibitemOpen
  \bibfield  {author} {\bibinfo {author} {\bibfnamefont {V.}~\bibnamefont
  {Vitagliano}}, \bibinfo {author} {\bibfnamefont {T.~P.}\ \bibnamefont
  {Sotiriou}}, \ and\ \bibinfo {author} {\bibfnamefont {S.}~\bibnamefont
  {Liberati}},\ }\href {\doibase 10.1103/PhysRevD.82.084007} {\bibfield
  {journal} {\bibinfo  {journal} {Phys.Rev.}\ }\textbf {\bibinfo {volume}
  {D82}},\ \bibinfo {pages} {084007} (\bibinfo {year} {2010})},\ \Eprint
  {http://arxiv.org/abs/1007.3937} {arXiv:1007.3937 [gr-qc]} \BibitemShut
  {NoStop}%
\bibitem [{\citenamefont {Sotiriou}\ and\ \citenamefont
  {Liberati}(2007)}]{Sotiriou:2006qn}%
  \BibitemOpen
  \bibfield  {author} {\bibinfo {author} {\bibfnamefont {T.~P.}\ \bibnamefont
  {Sotiriou}}\ and\ \bibinfo {author} {\bibfnamefont {S.}~\bibnamefont
  {Liberati}},\ }\href {\doibase 10.1016/j.aop.2006.06.002} {\bibfield
  {journal} {\bibinfo  {journal} {Annals Phys.}\ }\textbf {\bibinfo {volume}
  {322}},\ \bibinfo {pages} {935} (\bibinfo {year} {2007})},\ \Eprint
  {http://arxiv.org/abs/gr-qc/0604006} {arXiv:gr-qc/0604006 [gr-qc]}
  \BibitemShut {NoStop}%
\bibitem [{\citenamefont {Olmo}\ \emph {et~al.}(2009)\citenamefont {Olmo},
  \citenamefont {Sanchis-Alepuz},\ and\ \citenamefont
  {Tripathi}}]{Olmo:2009xy}%
  \BibitemOpen
  \bibfield  {author} {\bibinfo {author} {\bibfnamefont {G.~J.}\ \bibnamefont
  {Olmo}}, \bibinfo {author} {\bibfnamefont {H.}~\bibnamefont
  {Sanchis-Alepuz}}, \ and\ \bibinfo {author} {\bibfnamefont {S.}~\bibnamefont
  {Tripathi}},\ }\href {\doibase 10.1103/PhysRevD.80.024013} {\bibfield
  {journal} {\bibinfo  {journal} {Phys.Rev.}\ }\textbf {\bibinfo {volume}
  {D80}},\ \bibinfo {pages} {024013} (\bibinfo {year} {2009})},\ \Eprint
  {http://arxiv.org/abs/0907.2787} {arXiv:0907.2787 [gr-qc]} \BibitemShut
  {NoStop}%
\bibitem [{\citenamefont {Zel'dovich}\ and\ \citenamefont
  {Novikov}(1971)}]{ZeldovichNovikov}%
  \BibitemOpen
  \bibfield  {author} {\bibinfo {author} {\bibfnamefont {Y.~B.}\ \bibnamefont
  {Zel'dovich}}\ and\ \bibinfo {author} {\bibfnamefont {I.~D.}\ \bibnamefont
  {Novikov}},\ }\href@noop {} {\emph {\bibinfo {title} {{Relativistic
  Astrophysics, Vol. 1}}}}\ (\bibinfo {year} {1971})\ \bibinfo {note}
  {university of Chicago Press}\BibitemShut {NoStop}%
\bibitem [{\citenamefont {Shapiro}\ and\ \citenamefont
  {Teukolsky}(1983)}]{Shapiro:1983du}%
  \BibitemOpen
  \bibfield  {author} {\bibinfo {author} {\bibfnamefont {S.~L.}\ \bibnamefont
  {Shapiro}}\ and\ \bibinfo {author} {\bibfnamefont {S.~A.}\ \bibnamefont
  {Teukolsky}},\ }\href@noop {} {\emph {\bibinfo {title} {Black holes, white
  dwarfs, and neutron stars: the physics of compact objects}}}\ (\bibinfo
  {publisher} {John Wiley and Sons},\ \bibinfo {address} {New York},\ \bibinfo
  {year} {1983})\BibitemShut {NoStop}%
\bibitem [{\citenamefont {Kainulainen}\ \emph {et~al.}(2007)\citenamefont
  {Kainulainen}, \citenamefont {Piilonen}, \citenamefont {Reijonen},\ and\
  \citenamefont {Sunhede}}]{Kainulainen:2007bt}%
  \BibitemOpen
  \bibfield  {author} {\bibinfo {author} {\bibfnamefont {K.}~\bibnamefont
  {Kainulainen}}, \bibinfo {author} {\bibfnamefont {J.}~\bibnamefont
  {Piilonen}}, \bibinfo {author} {\bibfnamefont {V.}~\bibnamefont {Reijonen}},
  \ and\ \bibinfo {author} {\bibfnamefont {D.}~\bibnamefont {Sunhede}},\ }\href
  {\doibase 10.1103/PhysRevD.76.024020} {\bibfield  {journal} {\bibinfo
  {journal} {Phys.Rev.}\ }\textbf {\bibinfo {volume} {D76}},\ \bibinfo {pages}
  {024020} (\bibinfo {year} {2007})},\ \Eprint {http://arxiv.org/abs/0704.2729}
  {arXiv:0704.2729 [gr-qc]} \BibitemShut {NoStop}%
\bibitem [{\citenamefont {Olmo}(2008)}]{Olmo:2008pv}%
  \BibitemOpen
  \bibfield  {author} {\bibinfo {author} {\bibfnamefont {G.~J.}\ \bibnamefont
  {Olmo}},\ }\href {\doibase 10.1103/PhysRevD.78.104026} {\bibfield  {journal}
  {\bibinfo  {journal} {Phys.Rev.}\ }\textbf {\bibinfo {volume} {D78}},\
  \bibinfo {pages} {104026} (\bibinfo {year} {2008})},\ \Eprint
  {http://arxiv.org/abs/0810.3593} {arXiv:0810.3593 [gr-qc]} \BibitemShut
  {NoStop}%
\bibitem [{\citenamefont {Bertolami}\ \emph {et~al.}(2007)\citenamefont
  {Bertolami}, \citenamefont {Boehmer}, \citenamefont {Harko},\ and\
  \citenamefont {Lobo}}]{Bertolami:2007gv}%
  \BibitemOpen
  \bibfield  {author} {\bibinfo {author} {\bibfnamefont {O.}~\bibnamefont
  {Bertolami}}, \bibinfo {author} {\bibfnamefont {C.~G.}\ \bibnamefont
  {Boehmer}}, \bibinfo {author} {\bibfnamefont {T.}~\bibnamefont {Harko}}, \
  and\ \bibinfo {author} {\bibfnamefont {F.~S.}\ \bibnamefont {Lobo}},\ }\href
  {\doibase 10.1103/PhysRevD.75.104016} {\bibfield  {journal} {\bibinfo
  {journal} {Phys.Rev.}\ }\textbf {\bibinfo {volume} {D75}},\ \bibinfo {pages}
  {104016} (\bibinfo {year} {2007})},\ \Eprint {http://arxiv.org/abs/0704.1733}
  {arXiv:0704.1733 [gr-qc]} \BibitemShut {NoStop}%
\end{thebibliography}%
\end{document}